# Enhancement of polarizabilities of cylinders with cylinder-slab resonances

Meng Xiao[1], Xueqin Huang[1], H. Liu[2] and C. T. Chan[1*]

[1] Department of Physics and Institute for Advanced Study, the Hong Kong University of Science and Technology, Clear Water Bay, Hong Kong, China

[2] National Laboratory of Solid State Microstructures & Department of Physics, Nanjing University, Nanjing 210093, People's Republic of China

[*]Corresponding author: phchan@ust.hk

**Abstract:** If an object is very small in size compared with the wavelength of light, it does not scatter light efficiently. It is hence difficult to detect a very small object with light. We show using analytic theory as well as full wave numerical calculation that the effective polarizability of a small cylinder can be greatly enhanced by coupling it with a superlens type metamaterial slab. This kind of enhancement is not due to the individual resonance effect of the metamaterial slab, nor due to that of the object, but is caused by a collective resonant mode between the cylinder and the slab. We show that this type of particle-slab resonance which makes a small two-dimensional object much "brighter" is actually closely related to the reverse effect known in the literature as "cloaking by anomalous resonance" which can make a small cylinder undetectable. We also show that the enhancement of polarizability can lead to strongly enhanced electromagnetic forces that can be attractive or repulsive, depending on the material properties of the cylinder.

## Introduction

A metamaterial slab with both $\varepsilon$ and $\mu$ being negative can achieve negative refraction[1] that will in turn lead to many novel effects [2-5]. Such metamaterial slabs can be realized using an array of sub-wavelength resonators[6,7]. Of particular interest is a slab with $\varepsilon = \mu = -1$, which can image a point source with perfect resolution [2] and if absorption is taken into account, a slab with $\varepsilon = \mu = -1 + i\delta$ becomes a "superlens" that can beat the diffraction limit of ordinary lens[8]. This enhanced imaging capability is just one of the many enhanced effects attributed to metamaterials. Other examples include enhancing the performance of antennas[9] using metamaterial substrates (e.g. high impedance surfaces[10]), tuning reflection/transmission phases using meta-surfaces [11,12], and metamaterials can even be used to make objects invisible[13-15]. More examples can be found in recent review articles[16-18]. Metamaterial substrates are also known to induce enhancement or modification of optical responses due to field enhancement effect[19,20]. We note the "enhanced polarizability" has been considered in the context of molecules interacting with small clusters[21]. Here, we discuss a new kind of enhancement effect which strongly renormalizes the effective polarizability of a small 2D object.

The response of a small object to an external field is characterized by the polarizability α, which is typically proportional to the volume (3D) or area (2D). As such, we expect that a Rayleigh particle should have a small response, except in the case of plasmonic resonance condition is satisfied for the particle. It is difficult to detect or manipulate a small particle with light as it interacts weakly with external fields. This paper explores the possibility of enhancing the polarizability of a small object and hence its interaction with external fields by placing it in the proximity of a perfect lens type substrate. We are mostly interested in "off-particle-resonance" situations, referring to a polarizability enhancement due not to the intrinsic resonance of the particle (i.e. plasmon resonance condition is not satisfied), but to the resonant electromagnetic coupling of the particle with the substrate. The particle-substrate is also not due to the resonance condition of substrate, which will serve to suppress rather than enhance the effective polarizability of the cylinder. We will also explore whether such enhancements can induce some interesting physical effects such as strong electromagnetic forces.

The system we consider is shown schematically in Fig.1, where a cylinder with a very small radius is placed in front of a "superlens" type metamaterial slab that is infinitely extended in the x-y plane. We consider situations in which the bare scattering cross section of the cylinder is very small as the radius is much smaller than the incident wavelength. The radius of the cylinder, the distance from the center of the cylinder to the surface of the metamaterial slab and the thickness of the slab are given by $r_c$, $z_d$ and $d$, respectively. A plane wave impinges on the system with the electric (E) field along the y direction. The angle between the plane wave vector and the z axis is given by $\phi$.

In this paper, we use the normalized units which are more convenient for numerical simulations as all the qualities we want to compute and compare are expressed as ratios. To be more specific, we set $\varepsilon_0$ $\mu_0$ and the speed of light in a vacuum ($c$) to be unity. We set the thickness of the slab to be $d=1$ and all the other length scales are expressed in terms of $d$. We define a center frequency $\omega_c = 6\pi$ and at this frequency $\omega_c d/c = 6\pi$. The radius of the cylinder is chosen to be $r_c = 0.005d$, then $\omega_c r_c/c = 0.03\pi$ and hence the cylinder is small compared to the wavelength and the scattering should be in the Rayleigh regime as long as the relative permittivity of the cylinder ($\varepsilon_c$) is not too large. The cylinders studied in this paper are all assumed to be non-magnetic, i.e., $\mu_c = 1$. The relative permittivity ($\varepsilon_s$) and permeability ($\mu_s$) of the superlens metamaterial slab is given by

$$\varepsilon_s = \mu_s = 1 - \frac{\omega_{ps}^2}{\omega^2 + i\nu\omega}. \quad (1)$$

We set $\omega_{ps} = \sqrt{2}\omega_c, \nu = \frac{1}{2}10^{-6}\omega_c$, thus at $\omega = \omega_c$, $\varepsilon_s = \mu_s = -1 + 10^{-6}i$. For the calculation of the electromagnetic forces acting on the cylinder, we integrate the Maxwell Stress Tensor along the dashed circle (See Fig. 1) with a radius of $R = 1.5r_c$.

## Analytic approach

In the Rayleigh limit, the polarizability of the cylinder is small. We expect a stronger coupling between the incident wave and the cylinder if the cylinder is placed close to a reflector, as the reflector may serve to increase the field intensity near the cylinder. For the case of a metamaterial slab which carries surface waves, the excitation of the surface wave can lead to strong fields near the surface and hence the metamaterial slab should enhance the interaction between the cylinder and the external electromagnetic field. On the other hand, a superlens type metamaterial slab is designed to be almost impedance matched to air, except for the deviation from perfectness originating from the imaginary part of ε and μ. An almost impedance-matched interface should not offer much reflectance and if we argue heuristically in this manner, we do not expect much enhancement due to reflection. The question then arises: should a superlens type metamaterial slab enhance the polarizability of a small 2D particle? A more mathematical analysis is in order. To understand the possible enhancement effect, let us first seek an analytic solution which is feasible if we only consider the lowest order scattering of the small cylinder. In other words, the scattering of the cylinder is described by a monopole with polarizability given by[22]

$$\alpha = 4i\varepsilon_o \beta(r_c k_0)/k_0^2 , \quad (2)$$

where $k_0$ is the wave vector in the vacuum, and

$$\beta(r_c k_0) = \frac{J_0(n_c r_c k_0) J_0'(r_c k_0) - n_c J_0(r_c k_0) J_0'(n_c r_c k_0)}{J_0(n_c r_c k_0) H_0'^{(1)}(r_c k_0) - n_c H_0^{(1)}(r_c k_0) J_0'(n_c r_c k_0)} , \quad (3)$$

where $J_0$, $H_0^{(1)}$, $J_0'$ and $H_0'^{(1)}$ are the $0^{th}$ order of Bessel function, Hankel function of the first kind and their derivatives, $n_c = \sqrt{\varepsilon_c}$ is the refractive index of the cylinder. The monopole moment of the tiny cylinder along the y direction is given by

$$P_y = \alpha E_y^{loc} = \alpha \left( E_y^{inc} + \frac{k_0^2}{\varepsilon_0} W_{yy}^{ref} P_y \right) \quad (4)$$

where $E_y^{loc}$ and $E_y^{inc}$ are the y component of the local field and the incident electric field, respectively, $W_{yy}^{ref}$ is the yy component of the reflection part of the dyadic Green's function. In the second step, the local field is decomposed into two parts, which are the incident field (including the reflection from the metamaterial slab) and the monopole field reflected back by the metamaterial slab. After expanding in the plane wave basis, $W_{yy}^{ref}$ can be written as

$$W_{yy}^{ref} = \int_{-\infty}^{\infty} dk_p \frac{i}{4\pi k_z} \exp(2ik_z z_d) R(k_p), \quad (4)$$

where $R(k_p)$ is the reflection coefficient of a plane wave with the parallel component of the wave vector given by $k_p$, $k_z = \sqrt{k_0^2 - k_p^2}$ is the wave vector along the z direction. Equation. (4) can now be reformulated as

$$P_y = \frac{\alpha}{1 - k_0^2 \alpha W_{yy}^{ref} / \varepsilon_0} E_y^{inc} = \alpha^* E_y^{inc}, \quad (6)$$

where $\alpha^*$ is the effective polarizability of the cylinder. When the cylinder is small compared to the wave length, i.e., $k_0 r_c \ll 1$,

$$k_0^2 \alpha / \varepsilon_0 \approx \pi (\varepsilon_c - 1)(k_0 r_c)^2 \left[ 1 + i\pi (\varepsilon_c - 1)(k_0 r_c)^2 / 4 \right]. \quad (7)$$

When the cylinder is non-absorptive, $\text{Re}[\alpha] \sim (k_0 r_c)^2$ and $\text{Im}[\alpha] \sim (k_0 r_c)^4$, then $\text{Re}[\alpha] \gg \text{Im}[\alpha]$ when the cylinder is small.

If the metamaterial slab is a superlens with $\varepsilon_s = \mu_s = -1 + i\delta$ at frequency $\omega = \omega_c$ and $\delta$ is a small positive number, then in the limiting process $z_d / d \to 0$, $\omega \to \omega_c$ while still keeping $|1 - \omega/\omega_c| \gg \delta$, we have after some manipulations [23]

$$W_{yy}^{ref} \sim \frac{1}{\pi} \frac{\beta(1-\omega/\omega_c)+i\delta}{\beta^2(1-\omega/\omega_c)^2+\delta^2} \ln(d/z_d), (8)$$

where $\beta = \omega \partial \varepsilon_s / \partial \omega |_{\omega=\omega_c}$ is the slope of the relative permittivity of the slab at $\omega = \omega_c$. Thus the reflection Green function has a resonance feature around $\omega_c$ if $\delta \ll \beta|1-\omega/\omega_c|$, and

$$\text{Re}\left[W_{yy}^{ref}\right] \sim \frac{\ln(d/z_d)}{\beta(1-\omega/\omega_c)}, (9.a)$$

$$\text{Im}\left[W_{yy}^{ref}\right] \sim \frac{\ln(d/z_d)\delta}{\beta^2(1-\omega/\omega_c)^2}, (9b)$$

and the real part goes to infinity. In the limiting process $z_d/d \to 0$, $\omega \to \omega_c$, we also have $\left|\text{Re}\left[W_{yy}^{ref}\right]\right| \gg \left|\text{Im}\left[W_{yy}^{ref}\right]\right|$. Hence for a tiny cylinder, the condition of $k_o^2 \alpha W_{yy}^{ref}/\varepsilon_o \approx 1$ can always be satisfied at some combinations of $(z_d, \omega)$, at which $\alpha^* \gg \alpha$ and hence the effective scattering cross section of the cylinder is greatly enhanced. The maximum value of the enhancement is bounded by $1/\max\left\{\text{Im}[\varepsilon_c], \text{Re}|(\varepsilon_c-1)|(k_0 r_c)^2, \delta/|\beta(1-\omega/\omega_c)|\right\}$ up to some constants. From the methamatical analysis above, we can see that this kind of enhancement effect is not due to the individual effect of the metamaterial slab or the tiny cylinder. It is also not due to the resonance of the cylinder. It involves both the properties of the slab and the cylinder and as such, it is a cylinder-slab resonance. The frequency of this cylinder-slab resonance mode is higher (lower) than $\omega_c$ if $\varepsilon_c$ is smaller (larger) than 1.

We further note that at $\omega = \omega_c$,

$$W_{yy}^{ref} \sim \frac{i}{2\pi} \frac{d}{z_d} \left(\frac{2d}{z_d \delta}\right)^{1-2z_d/d} \left[\ln(\sqrt{d/z_d}\delta^{-1})\right]^{-1}, (10)$$

which diverges in the limiting process $z_d/d \to 0$ and $\delta \to 0$. So when $z_d/d$ or $\delta$ is small enough, according to Equation. (10), $\left|k_0^2 \alpha W_{yy}^{ref}/\varepsilon_0\right| \gg 1$ and $\left|\alpha^*/\alpha\right| \ll 1$. In other words, the effective polarizabiity of the cylinder becomes vanishing small. The cylinder is hence cloaked and becomes undetectable [13, 24-27]. This intriguing effect is known as "cloaking of two-dimensional polarizable discrete systems by anomalous resonance" and was first discovered by Milton and co-authors[28] in the quasi-static limit. The superlens slab is hence is rather peculiar system. While it is designed to image an object with ultra-high resolution, a superlens actually cloaks a small object near the surface if the operating frequency is precisely $\omega = \omega_c$ at which $\varepsilon_s, \mu_s$ is closest to -1. However, close to (but not at) $\omega = \omega_c$, there are some distances at which the small 2D object becomes very bright due to the enhanced polarizability and hence becomes very detectable.

To show the enhancement and cloaking effect, we show in Fig. 2(a) the relative effective polarizability of the monopole moment defined as $|\alpha^*/\alpha|$ as a function of $z_d$ and $\omega$ calculating using the analytic results. In this figure, we use a dielectric cylinder and the relative permittivity of this cylinder is $\varepsilon_c = 6$. We note that in the frequency range $\omega < \omega_c$, there is a very conspicuous red line which corresponds to large enhancement of the monopole polarizability due to the cylinder-slab collective resonance effect. While exactly at $\omega = \omega_c$, there is a sharp vertical blue line which shows the cloaking effect.

## Full wave simulations

The analytic results presented in the previous section considers only the monopole degree of freedom. As the field varies rapidly near the slab surface due to the cylinder-slab resonance, we check the results against a full wave calculation which includes the contribution of higher order (e.g. dipole) excitations. The scattering of a wave by a cylinder in the presence of a slab can be treated efficiently by expanding plane waves using cylindrical functions centered at the center of the cylinder. [See e.g. Borghi *et al*[29, 30]]. The field near the cylinder is expanded as a summation of cylindrical waves and by calculating the coefficients of the multipoles inside the cylinder, and then comparing them with those without the slab, we can obtain the enhancement factor for each multipole due to the cylinder-slab resonance. The fields everywhere can then be determined.

The full wave results show that the monopole degree of freedom in our system is still the dominating scattering term even when the cylinder is placed close to a metamaterial slab. For a standalone cylinder, the dipole component is very small compared with the monopole component to start with (dipole is smaller by a factor of ($\sim (k_0 r_c)^2$) than monopole), and numerical calculations show that the enhancement of the dipole component is only about 1.5 times that of the monopole component in our system., The amplitude of each multipole component in the presence of slab equals the amplitude without slab times the enhancement due to the slab ($1.5 \times (k_0 r_c)^2$ is still a small number) and so the monopole component still dominates in the presence of the metamaterial slab. To show the enhancement effect due to the cylinder-slab resonance using the full wave approach, we plot in Fig. 2(b) the ratio between the induced monopole moment of the cylinder with and without slab is plotted as a function of $z_d$ and $\omega$. We note that the scattering cross section is proportional to the square of the monopole moment and the enhancement will be even more conspicuous if we plotted the enhancement of cross section instead. The parameters used in this calculation are the same as those used in Fig. 2(a). Figures. 2(a) and 2(b) are nearly the same which shows that the monopole approximation employed in the analytic study is quite a good approximation.

In Figs. 2(a) and 2(b), we study the enhancement and cloaking effect of a dielectric cylinder ($\varepsilon > 0$) in front of the same metamaterial slab. We now continue to study the enhancement effect if the cylinder has $\varepsilon < 0$. As the frequency range of the cylinder-slab resonance mode is quite narrow, we can ignore the dispersion of the metallic cylinder and we also assume it to be lossless. In the following study, we set $\varepsilon_c = -4$. From Equations. (6)-(8), we know that the frequency range of the cylinder-slab resonance mode should now be higher than $\omega_c$. This analytic prediction is consistent with the numerical results shown Fig. 2(c). In this figure, we can see the cloaking effect (sharp vertical blue line) as well as a very conspicuous enhancement effect (red line) due to the cylinder-slab resonance. We also find some features that can be traced to the effect of the surface mode of the metamaterial slab which is marked with the black arrow in Fig. 2(c). The coupling of surface wave on either side of the metamatreial slab splits the surface wave into two modes, one above $\omega_c$ and one below $\omega_c$. It can be shown that the surface mode above $\omega_c$ always has a maximum frequency at which the group velocity along the slab is zero, and the density of states (DOS) is very high at that frequency. This high DOS effect also manifests itself in the effective polarization of the monopole moment.

## Cylinder-slab mode

To better understand the peculiar property of this cylinder-slab mode, we examine the field distribution. In Fig. 3(a), the amplitude of the electric field ($|E_y|$) near the cylinder is plotted. In this figure, we first consider the case where the plane wave incidents on the bare dielectric cylinder ($\varepsilon_c = 6$, $r_c = 0.005d$) when there is no metamaterial slab. The amplitude, the incident angle ($\phi$) and the frequency of the plane wave are set to be 1, 0 (normal incidence) and $0.99645\omega_c$. Note that the range of the color bar in Fig. 3(a) is quite narrow around 1, meaning that the incident field is minimally perturbed by the cylinder as the scattering of the tiny cylinder is extremely small. When we turn to the case where there is a metamaterial slab near the cylinder, the field distribution becomes very different. In Fig. 3(b), we show the $|E_y|$ near the small cylinder when the metamaterial slab is placed at $z_d = 0.02d$. The other parameters are the same those used in Fig. 3(a). From Fig. 2, we know that, when $\omega = 0.99645\omega_c$ and $z_d = 0.02d$, there exists a cylinder-slab resonance mode and the field intensity should be greatly enhanced. Fig. 3(b) shows exactly the same feature where the amplitude of $|E_y|$ is significantly higher than that in Fig. 3(a) [note the different scales used in Fig. 3a and Fig. 3b]. The strongest excited field is localized around the surface of the metamaterial slab right below the cylinder and decays away from the slab. The amplitude of $|E_y|$ is around 30 at the position of the cylinder, and this explain why the enhancement of the monopole moment is about 30 at the cylinder-slab resonance in Figs. 2(a) and 2(b). When there is only the metamaterial slab in the absence of the cylinder, the reflection from the slab is extremely small as a perfect lens is designed to be impedance matched to the environment. One can prove that, $\lim_{\delta \to 0} |R(k_p)| = 0$ [23] for $k_p < \omega/c$. In Fig. 3(c), we show the $\mathrm{Re}[E_y]$ in front of the slab when $\phi = \pi/4$. In this simulation, we just remove the cylinder and keep other parameters unchanged. The disturbance on the field distribution induced by the presence of the slab is unnoticeable. Figures 3(a)-3(c) together give strong evidence that the cylinder-slab resonance is not due to the individual effect of the metamaterial slab or the tiny cylinder; it must involve both the cylinder and the slab. In the above discussion, we focused on normal incidence. In Fig. 3(d), we study the amplitude of the electric field distribution for the system when the incident light is not normal to the slab by selecting $\phi = \pi/4$ while keeping all the other parameters unchanged. The field distribution seems quite similar to what we found for $\phi = 0$. The reflection of the plane wave from the slab is negligible, so the amplitude of the incident field ($\left|E_y^{inc}\right|$) at the position of the cylinder does not change much as the incident angle changes. The amplitude of the scattering field of the cylinder is the dominant component of the total field near the cylinder. It depends only on the effective polarizability of the cylinder-slab system, rather than the incident angle. This explains why the amplitude of the field distribution in Fig. 3(d) is similar to that in Fig. 3(b).

## Enhancement of electromagnetic force

In this section, we consider the enhancement of electromagnetic forces due to cylinder-slab resonance. When light is incident onto a small cylinder or small particle, the object will experience a light-induced force due to photon pressure, which is typically very small for a Rayleigh particle. When the object is placed close to a metamaterial surface, we expect a stronger force due to an enhancement of the effective polarizability. To calculate the electromagnetic force acting on the cylinder, we integrate the time-averaged Maxwell Stress Tensor along the dashed circle enclosing the cylinder as shown in Fig. 1[31-34]. We find that the enhancement is gigantic near the cylinder-slab resonance, and for that reason, we will present the results in the logarithmic scale. We define a function for the purpose of illustration:

$$g(x) = \begin{cases} \text{sgn}(x)\log_{10}(|x|), & |x| > 1 \\ 0, & |x| \leq 1 \end{cases}, \quad (11)$$

We consider the function $g(F_{Sz}/F_0)$, where $F_{Sz}$ and $F_0$ are the electromagnetic forces with and without the metamaterial slab, respectively. We only care about the region where the force is enhanced, and so we set $g(F_{Sz}/F_0) = 0$ if there is no enhancement (i.e. $|F_{Sz}/F_0| \leq 1$). The function keeps the sign of $F_{Sz}/F_0$, which gives the direction of the electromagnetic force and the log-scale makes the cylinder-slab resonance peaks easier for visual inspection. In Figs. 4(a) and 4(b), we show results where the cylinders have $\varepsilon > 0$ and $\varepsilon < 0$ respectively, and the parameters used are the same as those used in Figs. 2(a) and 2(c). The incident directions of the plane waves are normal to the metamaterial slab (positive z direction) in both cases. By symmetry, the only component of the electromagnetic force acting on the cylinder is along the z direction. Red (Blue) color in Fig. 4 indicates that the electromagnetic force is along the positive (negative) z direction and the electromagnetic force pushes (pulls) the cylinder. Alternatively, we may say that red color means that the cylinder is attracted towards the slab, while blue color means that the cylinder is repelled from the slab when an external plane wave illuminates the system normally from above the slab. This pushing force is greatly enhanced by the cylinder-slab resonance for dielectric $\varepsilon > 0$ cylinders when $\omega < \omega_c$ while the pulling force is strongly enhanced for $\varepsilon < 0$ cylinders when $\omega > \omega_c$. The amplitude of the force is more than 4 orders of magnitude higher than that without the metamaterial slab. The effect of the surface mode of the metamaterial slab can also be seen in Fig. 4(b) and marked with the black arrow. At the frequency $\omega = \omega_c$, the cylinder is "cloaked" when $z_d$ is small enough. In this case, the cylinder also cannot feel the electromagnetic force.

If the incident plane wave is not along the positive z direction, we break the mirror symmetry along the x direction and the cylinder should be subjected to an electromagnetic force along the x direction. As an example, we consider an incident angle of $\phi = \pi/4$ while keeping all the other parameters unchanged. A dielectric cylinder ($\varepsilon_c = 6$) is used in this simulation. In Figs. 5(a) and 5(b), we show the force along the z ($g(F_{Sz}/F_0)$) and the x

($g(F_{Sx}/F_0)$) direction, respectively. The force along the z direction is insensitive to the direction change of incident plane wave. It is consistent with the fact that the enhancement is a resonance effect and hence should not depend that much on way the resonance is excited. The enhancement of the force along the x direction is proportional to $\sin\phi |\alpha^*/\alpha|^2$ and it is much smaller than the enhancement of force along the z direction.

## Summary

In this paper, we found that a special kind of cylinder-slab resonance mode can strongly enhance the effective polarizability of a two-dimensional cylinder. This resonance is not due to the individual resonance of the cylinder and nor that of the slab. We give the condition of this hybrid resonance mode using an analytic model where we only consider the monopole moment of the cylinder. The results of the analytic model are consistent with those from full wave simulations. Our analytic result shows that this phenomena is closely related to the intriguing "cloaking by anomalous resonance" that is known in the literature. The "cloaking by anomalous resonance" phenomena is due to the resonance of the slab, which reduces the polarizability of the object to zero to make it invisible; while the collective cylinder-resonance does the opposite: it increases the polarizability of the particle to make it more visible. The mechanism of this cylinder-slab resonance mode is explored by examing of the field distribution. We also show that the cylinder-slab resonance mode can greatly enhance the electromagnetic force acting on the cylinder.

## Acknowledgments

This work is supported by Research Grant Council grant M-HKUST601/12 and by the Doctoral Program of Higher Education (20120091140005). We thank Dr. S. B. Wang for helpful discussions.

FIGURES

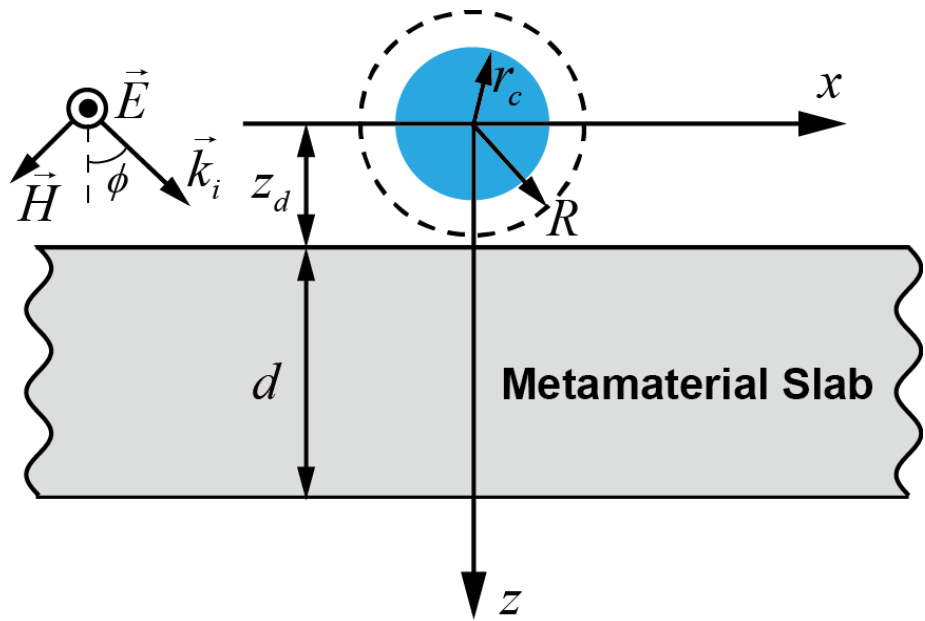

Fig. 1. Schematic picture of the system studied in this paper. A cylinder (blue solid circle) is placed in front of a matamaterial slab. The radius of the cylinder, the distance from the center of the cylinder to the surface of the metamaterial slab and the thickness of the metamaterial slab are given by $r_c$, $z_d$ and $d$, respectively. The electric field of the incident plane wave is along the y direction and the angle between the wave vector of the incident plane wave and the z axis is $\phi$. The electromagnetic force is calculated by integrating the Maxwell Stress Tensor along the dashed open circle around the cylinder.

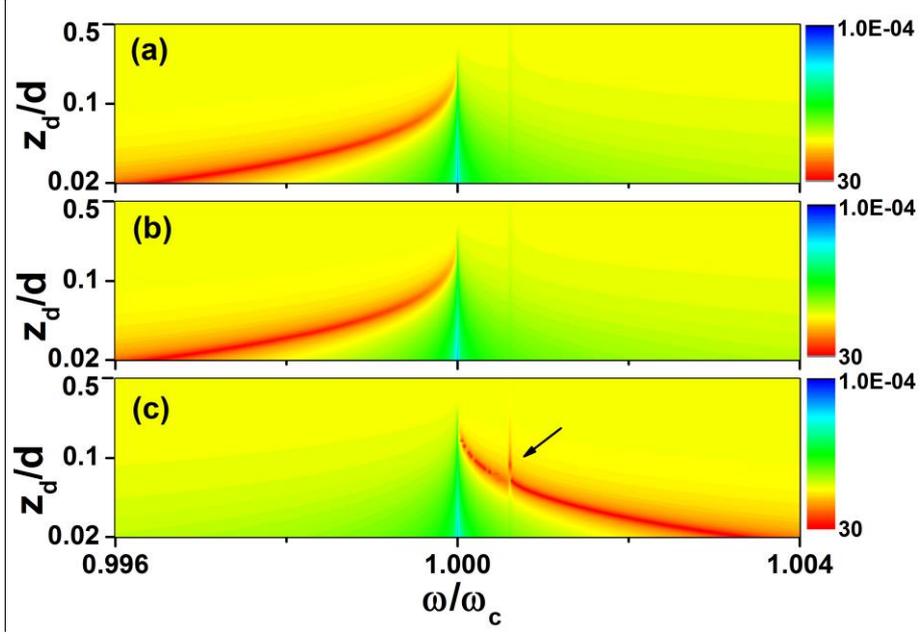

Fig. 2. The absolute value of the relative effective polarizability of monopole moment ($|\alpha^*/\alpha|$) as a function of $z_d$ and frequency $\omega$, where $\alpha$ is the polarizability of the monopole moment in the absence of the slab, and $\alpha^*$ is defined in the text. (a) Calculated using the analytic model for $\varepsilon_c = 6$, (b) Calculated using a full wave algorithm for $\varepsilon_c = 6$, (c) Calculated using full wave algorithm for $\varepsilon_c = -4$. The radius of the cylinder is $r_c = 0.005d$, where $d$ is the thickness of the metamaterial slab. The relative permittivity and permeability of the slab at $\omega_c$ are given by $\varepsilon_s(\omega_c) = \mu_s(\omega_c) = -1 + 10^{-6}i$. The blue vertical line at $\omega_c$ indicates that the relative effective polarizability is extremely small ("cloaking" effect). Red represents the region where the effective polarizability is greatly enhanced. The black arrow in (c) marks the frequency at which DOS of the surface wave of the slab is huge.

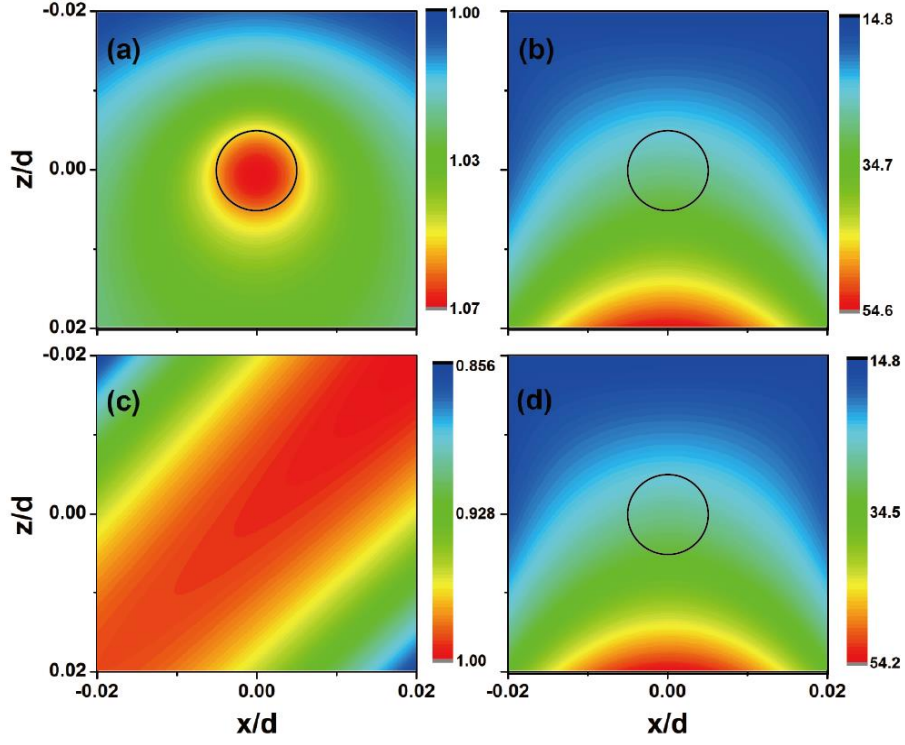

Fig. 3. (a), (c), (d) $|E_y|$ around the cylinder, which is marked by the black open circle. The relative permittivity, radius of cylinder are $\varepsilon_c = 6$ and $r_c = 0.005d$, respectively. (b) $\text{Re}[E_y]$ in front of the slab in the absence of the cylinder. There is no slab in (a), and $z_d = 0.02d$ in (b), (c) and (d). The plane wave propagates along the positive z direction in (a) and (b), and $\phi = \pi/4$ in (c) and (d). In (a)-(d), the frequency of the incident light is $\omega = 0.99645\omega_c$, and there is a cylinder-slab resonance mode at $z_d = 0.02d$ at this frequency.

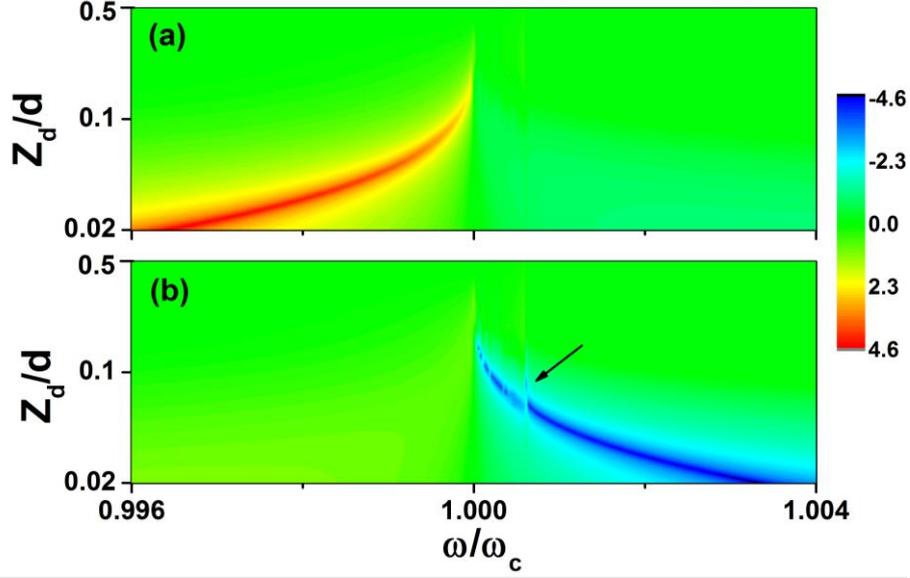

Fig. 4. $g\left(F_{Sz}/F_0\right)$ as a function of $z_d$ and $\omega$, where $F_{Sz}$ and $F_0$ are the electromagnetic force with and without the metamaterial slab, respectively, $g(x)=\mathrm{sgn}(x)\log_{10}(|x|)$ for $|x|\geq 1$ and $g(x)=0$ for $|x|<1$. The relative permittivity of the cylinder is $\varepsilon_c=6$ in (a) and $\varepsilon_c=-4$ in (b). In both (a) and (b), $r_c=0.005d$ and the direction of incident wave is along the positive z direction. The black arrow in (b) marks the frequency where DOS of the surface wave of the metamaterial slab is huge.

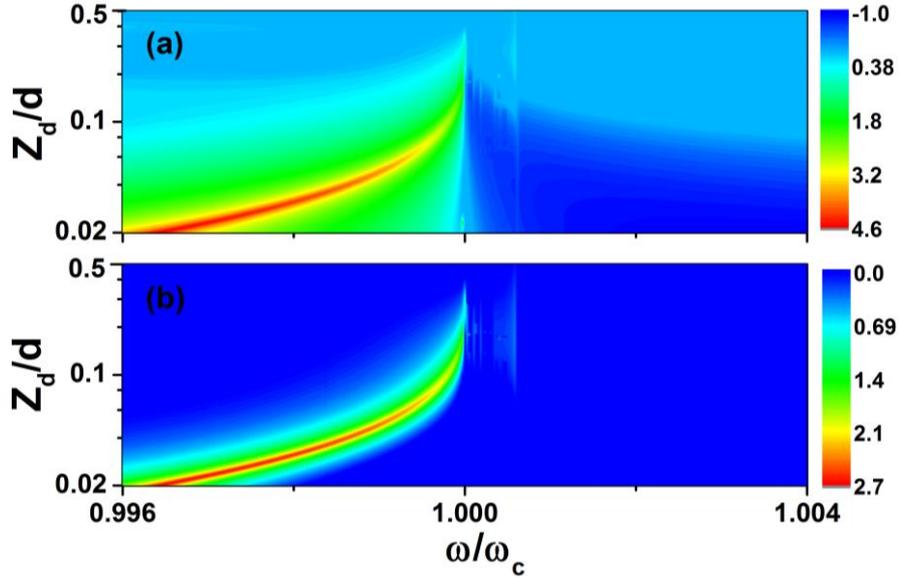

Fig. 5. (a) $g\left(F_{Sz}/F_0\right)$ and (b) $g\left(F_{Sx}/F_0\right)$ as a function of $z_d$ and frequency $\omega$, where $F_{Sz}$ ($F_{Sx}$) and $F_0$

are the electromagnetic force on the cylinder along the z (x) direction with the metamaterial slab and the force without the metamaterial slab, respectively, $g(x) = \text{sgn}(x)\log_{10}(|x|)$ for $|x| \geq 1$ and $g(x) = 0$ for $|x| < 1$. The relative permittivity, radius of the cylinders are $\varepsilon_c = 6$ and $r_c = 0.005d$, respectively. The angle between the wave vector of the incident wave and the positive z direction is $\phi = \pi/4$.